\documentclass[12pt,twoside,a4paper]{article}
\usepackage{amsmath,amssymb,latexsym,theorem,bbm,natbib,epsfig,color,subfigure}
\usepackage{multirow,graphicx,array,rotating}  
\usepackage{epstopdf,url,afterpage}
\newfont{\msa}{msam10 scaled\magstep1}
\newfont{\ssmsa}{msam9}

\def\crps{\mathop{\hbox{\rm CRPS}}}
\def\crpss{\mathop{\hbox{\rm CRPSS}}}

\numberwithin{equation}{section}

\title{Statistical post-processing of hydrological forecasts using Bayesian model averaging}

\author{{\sc S\'andor Baran$^{1}$},  {\sc Stephan Hemri$^{2}$} and {\sc Mehrez El Ayari$^{1}$} \vspace*{0.5cm}\\
         $^1$Faculty of Informatics, University of Debrecen\\
         Kassai \'ut 26, H-4028 Debrecen, Hungary \\
         $^2$Federal Office of Meteorology and Climatology MeteoSwiss\\ 
         Z\"urich-Airport, Switzerland
        }

\date{}
\begin{document}
\pagestyle{myheadings}

\maketitle

\begin{abstract}

Accurate and reliable probabilistic forecasts of hydrological quantities like runoff or water level are beneficial to various areas of society. Probabilistic state-of-the-art hydrological ensemble prediction models are usually driven with meteorological ensemble forecasts. Hence, biases and dispersion errors of the meteorological forecasts cascade down to the hydrological predictions and add to the errors of the hydrological models. The systematic parts of these errors can be reduced by applying statistical post-processing. For a sound estimation of predictive uncertainty and an optimal correction of systematic errors, statistical post-processing methods should be tailored to the particular forecast variable at hand. Former studies have shown that it can make sense to treat hydrological quantities as bounded variables. In this paper, a doubly truncated Bayesian model averaging (BMA) method, which allows for flexible post-processing of (multi-model) ensemble forecasts of water level, is introduced. A case study based on water level for a gauge of river Rhine, reveals a good predictive skill of doubly truncated BMA compared both to the raw ensemble and the reference ensemble model output statistics approach.

\end{abstract}

\section{Introduction}
\label{sec:sec1}

Hydrological forecasts are important for a heterogeneous group of users such as, for instance, the operators of hydrological power plants, flood prevention authorities, or shipping companies. For rational decision making based on cost-benefit analyses an estimate of the predictive uncertainty \citep{krzysztofowicz99, todini08} needs to be provided with any forecast. The state-of-the-art approach of using an ensemble of hydrological model runs driven by numerical weather prediction (NWP) models \citep{cp09} gives a first estimate of the meteorological input uncertainty. However, as NWP ensembles are usually biased and underdispersed \citep{bhtp,pbl08,btbbbc10} and other sources of uncertainty like hydrological model formulation, boundary and initial condition uncertainty as well as measurement uncertainties are typically neglected, statistical post-processing is important in order to reduce systematic errors and to obtain an appropriate estimate of the predictive uncertainty \citep{buizza18}. 

In the last decade various methods of statistical calibration of ensemble forecasts for different weather variables have been developed \citep[see e.g.][]{sk10,rs12,wfk} and some of them such as ensemble model output statistics \citep[EMOS;][]{grwg} or Bayesian model averaging \citep[BMA;][]{rgbp} provide full predictive distributions. The EMOS predictive distribution is given by a single parametric probability law with parameters depending on the ensemble, whereas the BMA predictive probability density function (PDF) is a weighted mixture of PDFs corresponding to the individual ensemble members. EMOS and BMA models for various weather quantities differ in the applied parametric distribution family and once the predictive distribution is given, its functionals (e.g. median or mean) can be considered as classical point forecasts. 

Besides the successful application e.g. to ensemble forecasts for temperature \citep{grwg}, wind speed \citep{tg,lt,bl15} or precipitation \citep{sch,schham,bn}, EMOS based statistical post-processing turned out to improve the predictive performance of hydrological ensemble forecasts for different gauges along river Rhine \citep{hlk15, hk17}. As EMOS is a quite parsimonious post-processing method that basically links a parametric forecast distribution to ensemble statistics like the ensemble mean and the ensemble variance, its performance is limited by i) how well the true process can be represented by a parametric distribution family and ii) to what extent the complete information from the ensemble can be summarized in a limited set of ensemble statistics. For instance, a typical EMOS approach based on a Gaussian or a Gamma distribution family is not able to model bimodal forecast distributions. However, BMA, which has also been applied to hydrological ensemble forecasts \citep[see e.g.][]{duan, hfz}, is much more flexible in that it converts a (multi-) model raw ensemble to a mixture distribution. Accordingly, we hypothesize that BMA may be able to outperform EMOS. However, the settings at river Rhine ask for double truncation in order to avoid physically unrealistic forecasts \citep{hk17}. To our best knowledge, up to now, there is no study that has applied a doubly truncated normal BMA approach. In this study, the work by \citet{bar}, which introduces a one-sided truncated normal BMA method, is extended to a two-sided truncated normal BMA approach. Its performance and its suitability for hydrological ensemble forecasts is assessed at the example of multi-model ensemble forecasts of water level at gauge Kaub at river Rhine. 

Doubly truncated BMA is introduced in Section \ref{sec:sec2} on calibration methods and forecast evaluation, followed by a brief description of the data in Section \ref{sec:sec3}. The results are presented in Section \ref{sec:sec4} and conclusions are drawn in Section \ref{sec:sec5}.

\section{Calibration methods and forecast evaluation}
\label{sec:sec2}

\subsection{Bayesian model averaging}
  \label{subs:subs2.1}
As mentioned in the Introduction, the BMA predictive distribution of a future weather quantity is a weighted mixture of probability laws corresponding to the individual ensemble members. In general, if \ $f_1,f_2, \ldots , f_K$ \ denote the ensemble forecast of a given weather or hydrological quantity \ $X$ \ for a given location, time and lead time, the BMA predictive PDF \citep{rgbp} of \ $X$ \ equals
\begin{equation}
  \label{eq:genBMA}
p(x\vert\, f_1, \ldots ,f_K;\theta_1, \ldots
,\theta_K):=\sum_{k=1}^K\omega _k g\big(x \vert\, f_k, \theta_k\big),
\end{equation}
where \ $g\big(x \vert f_k, \theta_k\big)$ \ is the component PDF from a parametric family corresponding to the $k$th ensemble member \ $f_k$ \ with parameter (vector) \ $\theta_k$ \ to be estimated and \ $\omega_k$ \ is the corresponding weight determined by the relative performance of this particular member during the training period. Note that the weights should form a probability distribution, that 
is \ $\omega_k\geq 0, \ k=1,2,\ldots ,K$ \ and \ $\sum_{k=1}^K \omega_k=1$. \ 

Recently most operational ensemble predictions systems (EPSs) incorporate ensembles where at least some members
can be considered as statistically indistinguishable and in this way exchangeable, as these forecasts are generated using perturbed initial conditions. This is the case with the 52 member operational ECMWF ensemble \citep{mbp,lp08} or one can mention multi-model EPSs such as the GLAMEPS ensemble \citep{iversen11} or the THORPEX Interactive Grand Global Ensemble \citep{tigge16}. Obviously, using exchangeable ensemble weather forecasts as inputs of a hydrological model results in hydrological ensemble forecasts with exchangeable members, which is the case with the water level data at hand described in Section \ref{sec:sec3}. To account for the existence of exchangeable ensemble groups \citet{frg2010} suggest to use the same weights and parameters within a given group. Thus, if we have \ $M$ \ ensemble members divided into \ $K$ \ exchangeable groups, where the \ $k$th \ group contains \ $M_k\geq 1$ \ ensemble members \ ($\sum_{k=1}^KM_k=M$) \ and \ $f_{k,\ell}$
\ denotes the  $\ell$th member of the $k$th group, model \eqref{eq:genBMA} is replaced by
\begin{equation}
  \label{eq:genBMAex}
p(x\vert f_{1,1},\ldots ,f_{1,M_1}, \ldots ,  f_{K,1},\ldots
,f_{K,M_K} ;\theta_1, \ldots, \theta_K):=\sum_{k=1}^K \sum_{\ell=1}^{M_k} \omega _k g\big(x \vert\, f_{k,\ell}, \theta_k\big). 
\end{equation}
For the sake of simplicity in the remaining part of this section we provide results and formulae only for model \eqref{eq:genBMA} as their extension to model \eqref{eq:genBMAex} is rather straightforward. Further, as in the case study of Section \ref{sec:sec4} the different lead times are treated separately, reference to the lead time is also omitted.

\subsection{Truncated normal BMA model}
\label{subs:subs2.2}

For weather variables such as temperature or pressure, BMA models with Gaussian components provide a reasonable fit \citep{rgbp,frg2010}, whereas wind speed calls for non-negative and skewed distributions such as gamma \citep{sgr10} or truncated normal with truncation from below at zero \citep{bar}. However, water levels are typically non-Gaussian \citep[see e.g.][]{duan}, moreover, they are bounded both from below and from above, which should also be taken into account during model formulation. A general procedure is to normalize the forecasts and observations using  Box-Cox transformation
\begin{equation}
  \label{eq:BC}
  h_{\lambda}(x):=\begin{cases} \big(x^{\lambda} -1\big)/\lambda, & \quad \lambda\ne 0, \\
  \log (x), & \quad \lambda =0 \end{cases}
\end{equation}
with some coefficient \ $\lambda$, \ perform post-processing, and then transform the results back using the inverse Box-Cox transformation \citep{duan,hlk14,hlk15}. Following the ideas of \citet{hk17}, for modelling Box-Cox transformed water levels we use a doubly truncated normal distribution \ ${\mathcal N}_a^b\big(\mu,\sigma^2 \big)$, \ with PDF
\begin{equation}
  \label{eq:tnormPDF}
g_{a,b}\big(x;\mu,\sigma\big):=\frac{\frac{1}{\sigma}\varphi\big(\frac{x-\mu}\sigma\big)}{\varPhi\big(\frac{b-\mu}\sigma\big)-\varPhi\big(\frac{a-\mu}\sigma\big)},\quad x\in [a,b],
\end{equation}
and \ $g_{a,b}\big(x;\mu,\sigma\big):=0$, \ otherwise, where \ $a$ \ and \ $b$ \ are the lower and upper bounds and \ $\varphi$ \ and \ $\varPhi$ \ denote the PDF and the cumulative distribution function (CDF) of the standard normal distribution, respectively. Note that the mean and variance of \ ${\mathcal N}_a^b\big(\mu,\sigma^2 \big)$ \ are
\begin{align}
  \label{eq:tnormMV}
  \kappa &= \mu + \sigma \frac{\varphi\big(\frac{a-\mu}\sigma\big) - \varphi \big(\frac{b-\mu}\sigma\big)}{\varPhi\big(\frac{b-\mu}\sigma\big)-\varPhi\big(\frac{a-\mu}\sigma\big)} \qquad \text{and}  \\  \varrho^2&=\sigma^2\left(1+ \frac{\frac{a-\mu}\sigma \varphi\big(\frac{a-\mu}\sigma\big) - \frac{b-\mu}\sigma \varphi \big(\frac{b-\mu}\sigma\big)}{\varPhi\big(\frac{b-\mu}\sigma\big)-\varPhi\big(\frac{a-\mu}\sigma\big)} - \Bigg(\frac{\varphi\big(\frac{a-\mu}\sigma\big) - \varphi \big(\frac{b-\mu}\sigma\big)}{\varPhi\big(\frac{b-\mu}\sigma\big)-\varPhi\big(\frac{a-\mu}\sigma\big)} \Bigg)^2 \right), \nonumber
\end{align}
respectively. The proposed  BMA predictive PDF is
\begin{equation}
   \label{eq:tnormBMA}
  p\big (x\mid f_1,\ldots ,f_K; \alpha_1, \ldots ,\alpha_K;\beta_1, \ldots ,\beta_K; \sigma \big)=\sum_{k=1}^K\omega_kg_{a,b}\big (x\mid\alpha_k+\beta_kf_k,\sigma\big),
\end{equation}  
where we assume that the location parameter of the $k$th mixture component is an affine function of the corresponding ensemble member $f_k$ and scale parameters are assumed to be equal for all component PDFs. The latter assumption is for the sake of simplicity and is common in BMA modelling \citep[see e.g.][]{rgbp}, whereas the form of the location parameter is in line with the truncated normal BMA model of \citet{bar}. Further, note that the EMOS model of \citet{hk17} also links location and scale of the truncated normal distribution to the ensemble members and not to the corresponding mean and variance. 

\subsection{Parameter estimation}
\label{subs:subs2.3}
Location parameters \ $\alpha_k, \ \beta_k$, \ weights \ $\omega_k, \ k=1,2,\ldots, M$, \ and scale parameter \ $\sigma$ \ can be estimated from training data, which consists of ensemble members and validating observations from the preceding \ $n$ \ days. In the BMA approach, estimates of location parameters are typically obtained by regressing the validating observations on the ensemble members, whereas weights and scale parameter(s) are obtained via maximum likelihood (ML) estimation \citep[see e.g.][]{rgbp,srgf,sgr10}, where the likelihood function of the training data is maximized using the EM algorithm for mixture distributions \citep{dlr,mclk}. However, the regression approach assumes the location parameters to be simple functions of the mean, which is obviously not the case for the truncated normal distribution. Hence, we propose a pure ML method, which ideas have already been considered e.g. by \citet{sgr10} or \citet{bar}.

In what follows, for a given location \ $s\in{\mathcal S}$ \ and  time \ $t\in{\mathcal T}$ \ let \ $f_{k,s,t}$ \ denote the $k$th ensemble member, and denote by \ $x_{s,t}$ \ the corresponding validating observation. By  assuming the conditional independence of forecast errors with respect to the ensemble members in space and time, the log-likelihood function for model \eqref{eq:tnormBMA} corresponding to all forecast cases \ $(s,t)$ \ in the training set equals
\begin{equation}
  \label{eq:logLik}
\ell (\omega_1,\ldots ,\omega_K,  \alpha_1, \ldots ,\alpha_K ,\beta_1, \ldots ,\beta_K, \sigma)=\sum_{s,t}\log \left[
  \sum_{k=1}^K \omega_k g_{a,b}\big(x_{s,t}\vert \, \alpha_k+\beta_k f_{k,s,t},
    \sigma \big) \right].
\end{equation}
To obtain the ML estimates we apply EM algorithm for truncated Gaussian mixtures proposed by \citet{ls12} with a mean correction. In line with the classical EM algorithm for mixtures \citep{mclk}, first we introduce latent binary indicator variables \ $z_{k,s,t}$ \ identifying the mixture component where the observation \ $x_{s,t}$ \ comes from. Using these indicator variables one can provide the  complete data log-likelihood corresponding to \eqref{eq:logLik} in the form
\begin{align}
  \label{eq:ClogLik}
  \ell_C (\omega_1,\ldots ,\omega_K, \alpha_1, \ldots ,\alpha_K ,&\,\beta_1, \ldots ,\beta_K,  \sigma)\\ 
  &=\sum_{s,t}\sum_{k=1}^Kz_{k,s,t} \left[ \log \big(\omega_k\big) + \log \Big(g_{a,b}\big(x_{s,t}\vert \, \mu_{k,s,t},
    \sigma \big)\Big) \right], \nonumber
\end{align}
with \ $\mu_{k,s,t}:=\alpha_k+\beta_k f_{k,s,t}$. \ After specifying the initial values of the parameters the EM algorithm alternates between an expectation (E) and a maximization (M) step until convergence. As first guesses \ $a_k^{(0)}$ \ and \ $b_k^{(0)}, \ k=1,2, \ldots, K$, \ for the location parameters one can use the coefficients of the linear regression of \ $x_{s,t}$ \ on  \ $f_{k,s,t}$, \ so \ $\mu^{(0)}_{k,s,t}=\alpha^{(0)}_k+\beta^{(0)}_k f_{k,s,t}$, \ initial scale \ $\sigma^{(0)}$ \  can be the standard deviation of the observations in the training data set, whereas the initial weights might be chosen uniformly, that is \ $\omega_k^{(0)}=1/K, \ k=1,2,\ldots ,K$. \ Then
in the E step the latent variables are estimated using the conditional expectation of the complete log-likelihood on the observed data, while in the M step the parameter estimates are updated by maximizing \ $\ell_C$ \ given the actual values of the latent variables.

For the doubly truncated normal model specified by \eqref{eq:tnormPDF} and \eqref{eq:tnormBMA} the E step of the $(j+1)$st iteration is
\begin{equation}
  \label{eq:stepE}
z_{k,s,t}^{(j+1)}:=\frac {\omega_k^{(j)}g_{a,b}\big(x_{s,t}\vert \,
  \mu_{k,s,t}^{(j)}, \sigma^{(j)} \big)}{\sum
  _{i=1}^K\omega_i^{(j)}g_{a,b}\big(x_{s,t}\vert \, 
  \mu_{i,s,t}^{(j)}, \sigma^{(j)} \big)}.
\end{equation}
Once the estimates of the indicator variables (which are not necessary $0$ or $1$ any more) are given, the first part of the M step updating the weights is obviously
\begin{equation}
  \label{eq:stepM1}
\omega_k^{(j+1)}:=\frac 1N\sum_{s,t}z_{k,s,t}^{(j+1)}, 
\end{equation}
where \ $N$ \ is the total number of forecast cases in the training set.

Further, non-linear equations \ $\frac {\partial\ell_C}{\partial \alpha_k}=0$ \ and \ $\frac {\partial\ell_C}{\partial \beta_k}=0, \ k=1,2, \ldots ,K$, \
lead us to update formulae
\begin{align}
  \label{eq:updateAB}  \alpha_{k}^{(j+1)}\!&:=\!\left[\sum_{s,t}z_{k,s,t}^{(j+1)}\right]^{-1}\!\!\!\sum_{s,t} z_{k,s,t}^{(j+1)}\left\{ \big(x_{k,s,t}-\beta_{k,s,t}^{(j)}f_{k,s,t}\big )+\sigma^{(j)}\frac{\varphi\Big(\frac{b-\mu_{k,s,t}^{(j)}}{\sigma^{(j)}}\Big)\!-\! \varphi\Big(\frac{a-\mu_{k,s,t}^{(j)}}{\sigma^{(j)}}\Big)}{\varPhi\Big( \frac{b-\mu_{k,s,t}^{(j)}}{\sigma^{(j)}}\Big)\!-\!\varPhi\Big(\frac{a-\mu_{k,s,t}^{(j)}}{\sigma^{(j)}}\Big)}\right\},\\   \beta_{k}^{(j+1)}\!&:=\!\left[\sum_{s,t}z_{k,s,t}^{(j+1)}f_{k,s,t}^{2}\right]^{-1}\!\!\!  \sum_{s,t}z_{k,s,t}^{(j+1)}f_{k,s,t}\left\{ \big(x_{k,s,t}\!-\!\alpha_{k,s,t}^{(j)}\big)\!+\!\sigma^{(j)}\frac{ \varphi\Big(\frac{b-\mu_{k,s,t}^{(j)}}{\sigma^{(j)}}\Big)\!-\!\varphi \Big(\frac{a-\mu_{k,s,t}^{(j)}}{\sigma^{(j)}}\Big)}{\varPhi\Big( \frac{b-\mu_{k,s,t}^{(j)}}{\sigma^{(j)}}\Big)\!-\!\varPhi\Big(\frac{a-\mu_{k,s,t}^{(j)}}{\sigma^{(j)}}\Big)}\right\}, \nonumber
\end{align}
respectively. However, using then simply \  $\mu_{k,s,t}^{(j+1)}:=\alpha_k^{(j+1)}+\beta_k^{(j+1)} f_{k,s,t}$ \ as the update of the location parameter in our case study results in an unstable parameter estimation process, so similar to \citet{bar} we introduce a mean correction of form
\begin{equation}
\label{eq:updateMu}  \mu_{k,s,t}^{(j+1)}:=\mu_{k,s,t}^{(0)}-\sigma^{(j)}\frac{\varphi\Big(\frac{a-\alpha^{(j+1)}-\beta^{(j+1)}f_{k,s,t}}{\sigma^{(j)}}\Big)-\varphi\Big( \frac{b-\alpha^{(j+1)}-\beta^{(j+1)}f_{k,s,t}}{\sigma^{(j)}}\Big)}{\varPhi \Big(\frac{b-\alpha^{(j+1)}-\beta^{(j+1)}f_{k,s,t}}{\sigma^{(j)}}\Big)-\varPhi\Big( \frac{a-\alpha^{(j+1)}-\beta^{(j+1)}f_{k,s,t}}{\sigma^{(j)}}\Big)},
\end{equation}
which reflects to the difference between the location and mean of a truncated normal distributions, see \eqref{eq:tnormMV}. Finally, from \ $ \frac {\partial\ell_C}{\partial \sigma}=0$ \ we obtain the last update formula
\begin{align}
  \label{eq:updateSigma}
   \sigma^{2(j+1)}:=\frac 1N\sum_{s,t}\sum_{k=1}^Kz_{k,s,t}^{(j+1)}\Bigg\{ &\big(x_{s,t}-\mu_{k,s,t}^{(j+1)}\big)^{2} \\      &+\left . \sigma^{(j)}\frac{\big(b-\mu_{k,s,t}^{(j+1)}\big)\varphi\Big(\frac{b-\mu_{k,s,t}^{(j+1)}}{\sigma^{(j)}}\Big)\!-\!\big(a-\mu_{k,s,t}^{(j+1)}\big)\varphi\Big(\frac{a-\mu_{k,s,t}^{(j+1)}}{\sigma^{(j)}}\Big)}{\varPhi\Big(\frac{b-\mu_{k,s,t}^{(j+1)}}{\sigma^{(j)}}\Big)\!-\!\varPhi\Big(\frac{a-\mu_{k,s,t}^{(j+1)}}{\sigma^{(j)}}\Big)}\right\}. \nonumber
\end{align}
Note that without truncation \ ($-a=b=\infty$) \ the terms of \eqref{eq:updateAB} and \eqref{eq:updateSigma} depending on \ $\sigma^{(j)}$ \ disappear, so location (mean) and scale (standard deviation) are updated separately, no mean correction is required, and we get back the classical EM algorithm for normal mixtures.

As a more simple alternative approach one can omit the update step \eqref{eq:updateAB} for \ $\alpha_k$ \ and \ $\beta_k$, \ modify the mean correction step \eqref{eq:updateMu} to
\begin{equation}
  \label{eq:updateMuS}  \mu_{k,s,t}^{(j+1)}:=\mu_{k,s,t}^{(0)}-\sigma^{(j)}\frac{\varphi\Big(\frac{a-\mu_{k,s,t}^{(j)}}{\sigma^{(j)}}\Big)-\varphi\Big( \frac{b-\mu_{k,s,t}^{(j)}}{\sigma^{(j)}}\Big)}{\varPhi \Big(\frac{b-\mu_{k,s,t}^{(j)}}{\sigma^{(j)}}\Big)-\varPhi\Big( \frac{a-\mu_{k,s,t}^{(j)}}{\sigma^{(j)}}\Big)},
\end{equation}
and after the EM algorithm stops, estimate location parameters \ $\alpha_k$ \ and \ $\beta_k$ \ from a linear regression of the final value of \ $\mu_{k,s,t}$ \ on \ $f_{k,s,t}$.

Finally, one can also try the classical naive approach, where  location parameters \ $\alpha_k$ \ and \ $\beta_k$ \ are not updated at all, that is \
$\mu_{k,s,t}^{(j+1)}\equiv \alpha^{(0)}_k+\beta^{(0)}_k f_{k,s,t}$. \

In the case study of Section \ref{sec:sec4} the latter two approaches do not show significantly different forecast skills in terms of the verification scores defined in Section \ref{subs:subs2.4}, so only the results for the naive and pure ML approaches are reported.

\subsection{Verification scores}
\label{subs:subs2.4}

In probabilistic forecasting the principal aim is to access the maximal sharpness of the predictive distribution subject to calibration \citep{gbr}, where the former means a statistical consistency between the predictive distributions and the validating observations, whereas the latter refers to the concentration of the predictive distribution. One of the simplest tools for getting a first impression about the calibration of forecast distributions is the probability integral transform (PIT) histogram. By definition, the PIT is the value of predictive CDF at the validating observation \citep{rgbp}, which in case of proper calibration should follow a uniform distribution on the $[0,1]$ interval. In this way the PIT histogram is the continuous counterpart of the verification rank histogram for the raw ensemble defined as histogram of ranks of validating observations with respect to the corresponding ensemble forecasts \citep[see e.g.][Section 7.7.2]{wilks}. Again, for a properly calibrated ensemble the ranks should be uniformly distributed.

Predictive performance can be quantified with the help of scoring rules, which are loss functions assigning numerical values to pairs of forecasts and observations. In hydrology and atmospheric sciences one of the most popular scoring rules is the continuous ranked probability score \citep[CRPS;][]{grjasa,wilks}, as it assesses calibration and sharpness simultaneously. For a (predictive) CDF \ $F(y)$ \ and real value (observation) \ $x$ \ the CRPS is defined as
\begin{align}
  \label{eq:CRPS}
\crps\big(F,x\big):=\int_{-\infty}^{\infty}\big (F(y)-{\mathbbm 
  1}_{\{y \geq x\}}\big )^2{\mathrm d}y&=\int_{-\infty}^x F^2(y){\mathrm d} y +\int_x^{\infty}\big (1-F(y)\big )^2{\mathrm d}y \\
&={\mathsf E}|X-x|-\frac 12
{\mathsf E}|X-X'|, \nonumber
\end{align}
where \ ${\mathbbm 1}_H$ \ denotes the indicator of a set \ $H$, \ whereas \ $X$ \ and \ $X'$ \ are independent random variables with CDF \ $F$ \ and finite first moment. CRPS is a negatively oriented proper scoring rule \citep{grjasa}, that is the smaller the better, and the right-hand side of \eqref{eq:CRPS} shows that it can be expressed in the same unit as the observation. For truncated normal distribution the CRPS has a simple closed form \citep[see e.g. the {\tt R} package {\tt scoringRules};][]{jkl}, whereas for the truncated normal mixture \eqref{eq:tnormBMA} the second integral expression in the definition \eqref{eq:CRPS} should be evaluated numerically. Moreover, in our case study each calibration approach provides a predictive CDF \ $F$ \ for the Box-Cox transformed water level \ $X \in [a,b]$. \ Thus, the CRPS corresponding to the predictive CDF \ $G(y):=F\big(h_{\lambda }(y)\big)$ \ of the original water level \ $Y = h_{\lambda}^{-1}(X) \in \big[h_{\lambda}^{-1}(a),h_{\lambda}^{-1}(b)\big]$ \ and a real value \ $y$ \ equals
\begin{equation}
  \label{eq:bcCRPS}
 \crps\big(G,y\big) =\int_{h_{\lambda}^{-1}(a)}^y F^2\big(h_{\lambda}(u)\big){\mathrm d} u +\int_y^{h_{\lambda}^{-1}(b)}\Big (1-F\big (h_{\lambda}(u)\big)\Big )^2{\mathrm d}u,
\end{equation}
which integral should again be approximated numerically. Further, in order to get the CRPS of the raw ensemble, the predictive CDF should be replaced by the empirical one.

One can quantify the improvement in CRPS with respect to a reference predictive distribution \ $F_{ref}$ \ with the help of the continuous ranked probability skill score \citep[CRPSS;][]{murphy73,grjasa}, defined as
\begin{equation*}
\crpss \big(F,x;F_{ref}\big):=1-\frac{\crps \big(F,x\big)}{\crps \big(F_{ref},x\big)}.
\end{equation*}
In contrast to the CRPS the CRPSS is positively oriented, that is the larger the better, and in our case study we consider the raw ensemble as a reference.

Calibration and sharpness of a predictive distribution can also be investigated using the coverage and average width of the \ $(1-\alpha )100\,\%, \ \alpha \in (0,1),$ \ central prediction interval, respectively. As coverage we consider the proportion of validating observations located between the lower and upper \ $\alpha /2$ \  quantiles of the predictive CDF, and level
\ $\alpha$ \ should be chosen to match the nominal coverage of the raw ensemble, that is \ $(K-1)/(K+1)100\%$, \ where \ $K$ \ is the ensemble size.  As the coverage of a calibrated predictive distribution should be around \ $(1-\alpha )100\,\%$, \ such a choice of \ $\alpha$ \ allows direct comparison with the raw ensemble.

Further, as point forecasts we consider the medians of the predictive distributions and the raw ensemble, that are evaluated with the help of mean absolute errors (MAEs).

Finally, as suggested by \citet{gr11}, statistical significance of the differences between the verification scores is assessed by utilizing the Diebold-Mariano \citep[DM;][]{dm95} test, which allows accounting for the temporal dependencies in the forecast errors.

\subsection{Truncated normal EMOS model}
\label{subs:subs2.5}
As a reference post-processing method for calibration of Box-Cox transformed ensemble forecasts for water levels we consider the truncated normal EMOS model of \citet{hk17}. In this approach the predictive distribution is a single doubly truncated normal distribution \ $\mathcal N_a^b\big(\mu,\sigma ^2\big)$ \ defined by \eqref{eq:tnormPDF}, and the ensemble members are just linked to the location \ $\mu$ \ and scale \ $\sigma$ \ via equations
\begin{equation}
   \label{eq:emos}
 \mu = a_0+a_1f_1+ \cdots +a_Kf_K \qquad \text{and} \qquad  \sigma^2 = b_0+b_1 S^2, 
\end{equation}
where \ $S^2$ \ denotes the variance of the transformed ensemble. In case of existence of groups of exchangeable ensemble members the equation for the location in \eqref{eq:emos} is replaced by
\begin{equation}
  \label{eq:emosEx}
  \mu = a_0+a_1\overline f_1+ \cdots +a_K\overline f_K,
\end{equation}
where \ $\overline f_k$ \ denotes the mean value of the $k$th group.
According to the optimum score estimation principle of \citet{grjasa}, location parameters \ $a_0,a_1, \ldots ,a_K \in {\mathbb R}$ \ and scale parameters \ $b_0, b_1 \geq  0$ \ are estimated from the training data
by optimizing a proper verification score, which is usually the CRPS defined by \eqref{eq:CRPS}.

\section{Data}
\label{sec:sec3}
BMA and EMOS calibration approaches are tested on ensemble forecasts for water level (cm) at gauge Kaub of river Rhine (546 km) for the eight year period 1 January 2008 -- 31 December 2015 with lead times from 1 h to 120 h with a time step of 1 h and the corresponding validating observations. The minimum and maximum recorded water levels at this particular gauge are 35 cm and 825 cm, respectively. Our 79 member multimodel water level ensemble is obtained by plugging the ECMWF high resolution (HRES) forecasts, the 51 member ECMWF forecasts (EPS) \citep{mbp,lp08}, the 16 member COSMO LEPS forecasts of the limited-area ensemble prediction system of the consortium for small-scale modeling \citep{mcmp} and the 11 member NCEP GEFS forecasts of the reforecast version 2 of the global ensemble forecast system of the National Center for Environmental Prediction \citep{hamill13} for there relevant weather variables into the hydrological model HBV-96 \citep{ljpgb}, which is run at the German Federal Institute of Hydrology (BfG) for operational runoff forecasting. The runoff forecasts are then converted into water level forecasts for the navigation-relevant gauges, including gauge Kaub, using a hydrodynamic model. All ensemble forecast are initialized at 6 UTC. We remark that the data set at hand is part of the data studied in \citet{hk17}, where we refer to for further details on the data.

\section{Results}
\label{sec:sec4}

\begin{figure}[t]
  \begin{center}
    \epsfig{file=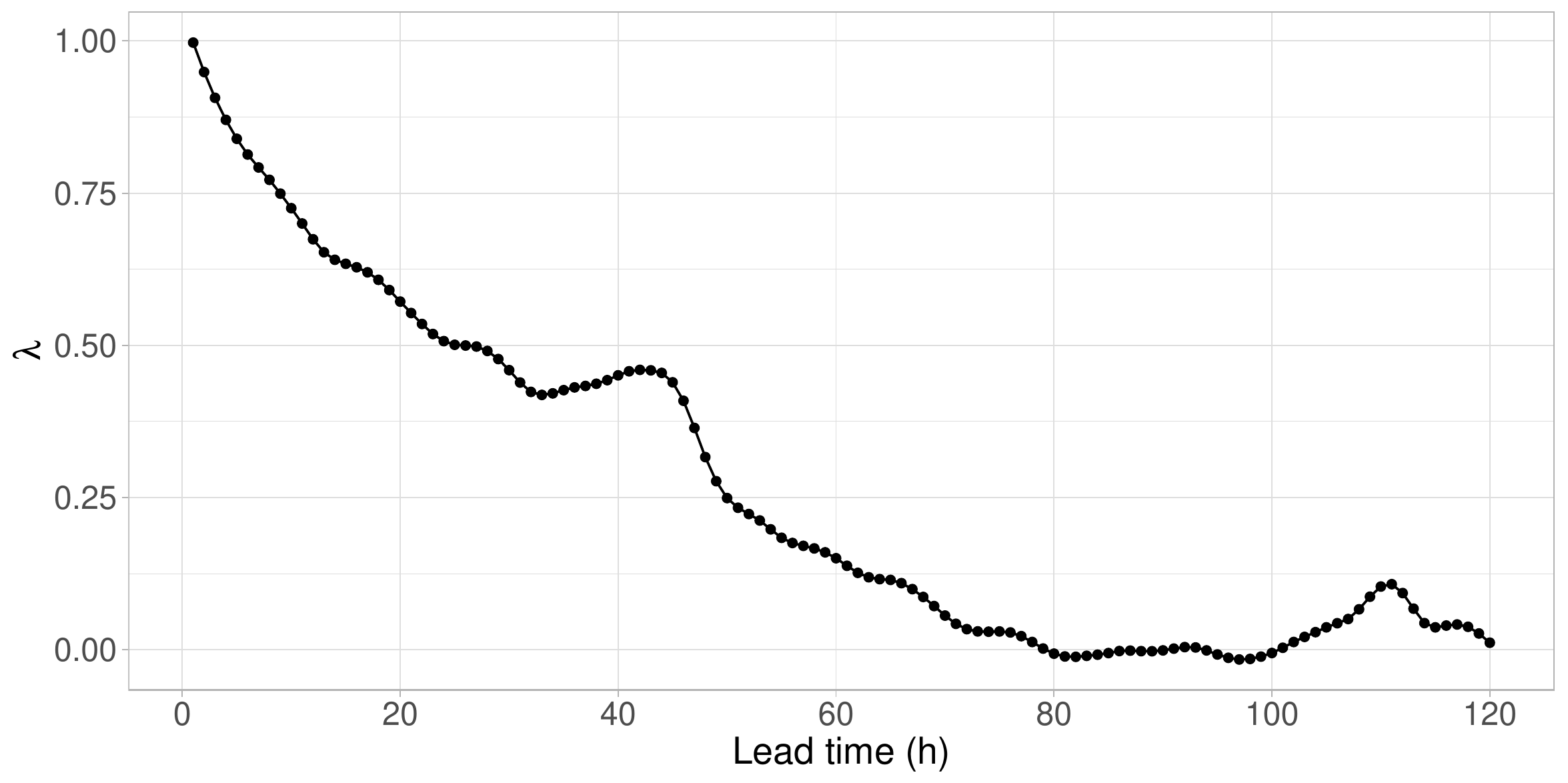, width=.7\textwidth}
    \end{center}
\caption{Box-Cox transformation parameter \ $\lambda$ \ as function of the lead time.}
\label{fig:BCcoeffs}
\end{figure}

As mentioned in Sections \ref{subs:subs2.2} and \ref{subs:subs2.5}, BMA and EMOS post-processing is applied for modelling Box-Cox transformed water levels.
Each lead time has an individual Box-Cox parameter \ $\lambda$ \ (see Figure \ref{fig:BCcoeffs}) providing the best fit to a normal distribution \citep{hlk15,hk17} and obviously, for a given lead time the same coefficient is applied both for the forecasts and observations.

\begin{figure}[!t]
    \epsfig{file=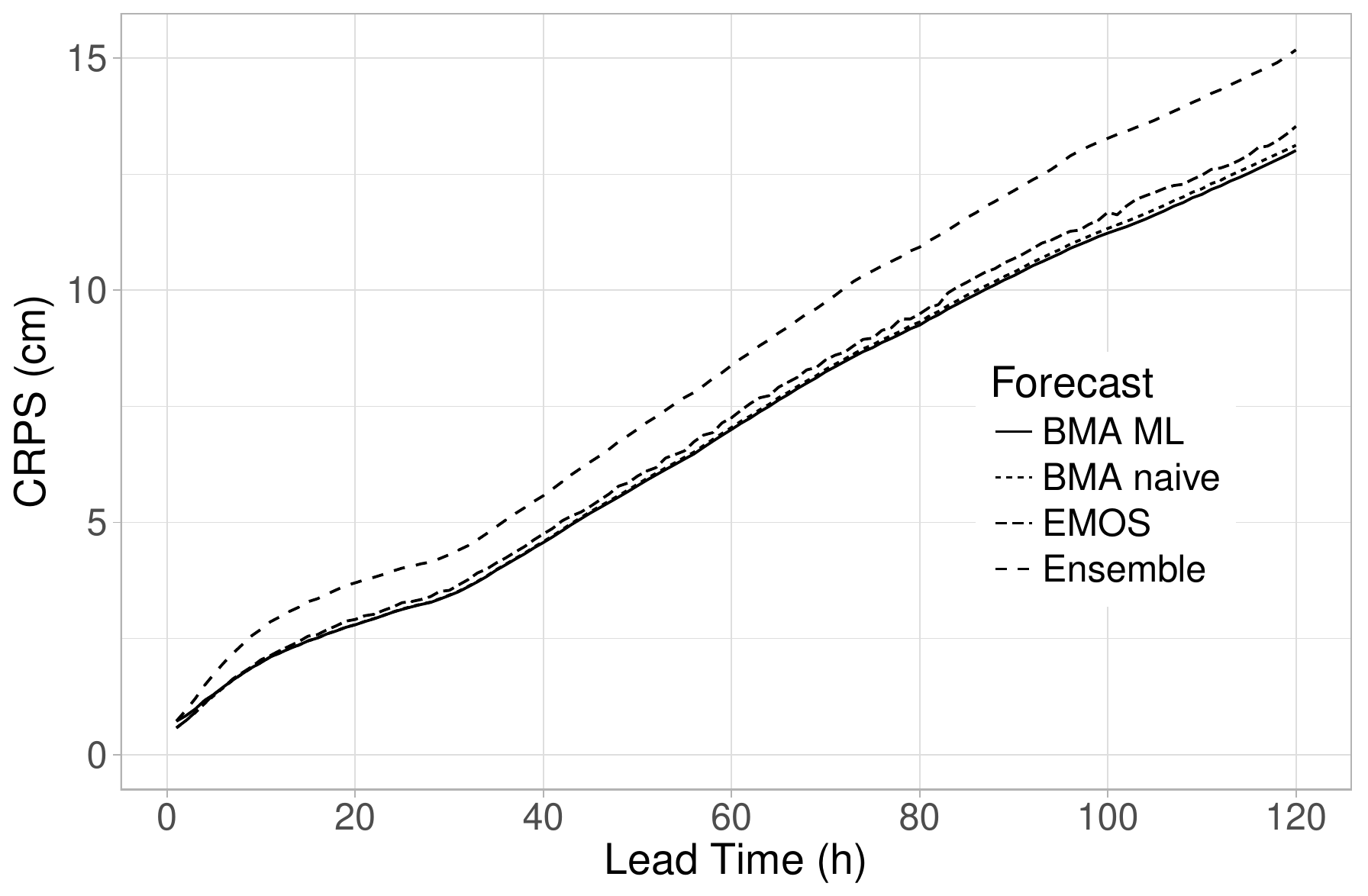, width=.49\textwidth} \ \
    \epsfig{file=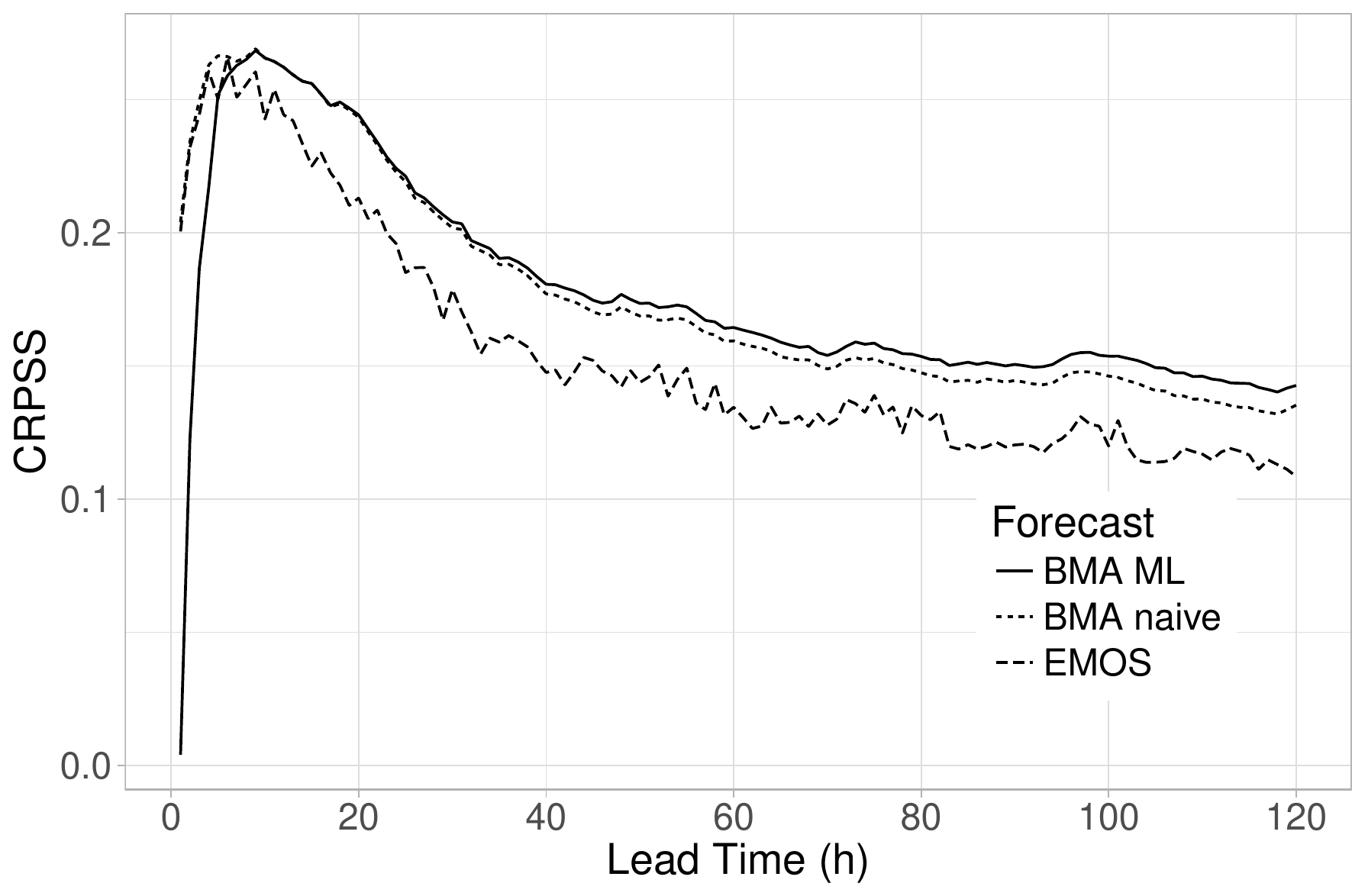, width=.49\textwidth}
    \centerline{ \hbox to 8.8 truecm{\scriptsize (a) \hfill (b)}}

    \bigskip
    \epsfig{file=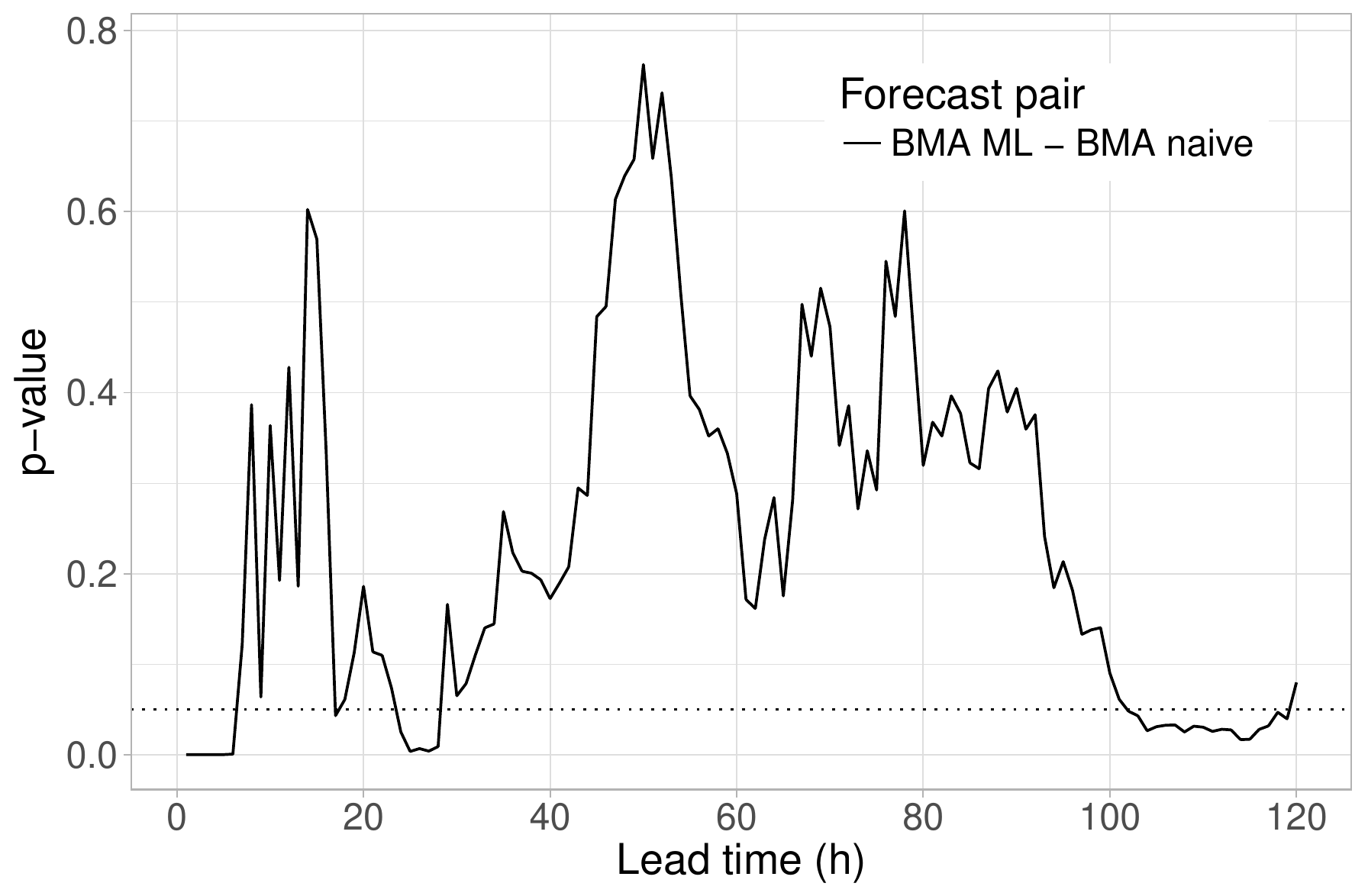, width=.49\textwidth} \ \
    \epsfig{file=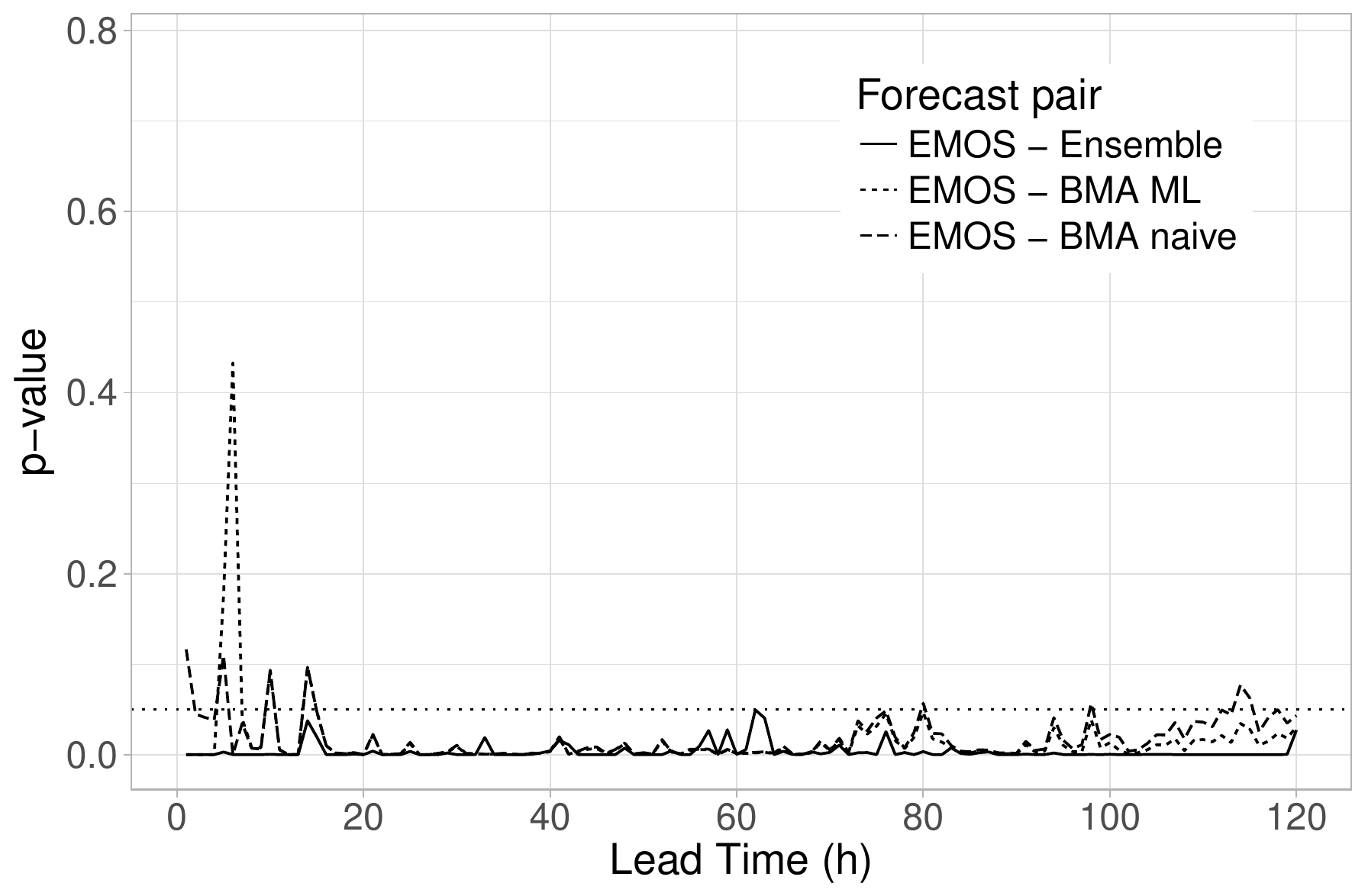, width=.49\textwidth}
   \centerline{ \hbox to 8.8 truecm{\scriptsize (c) \hfill (d)}}
\caption{Mean CRPS values (a) and CRPSS with respect to the raw ensemble (b); $p$-values of DM tests for equality of mean CRPS of the two BMA approaches (c) and of all models compared to EMOS (d). Horizontal dotted lines of (c) and (d) indicate a $5\,\%$ level of significance.}
\label{fig:crps}
\end{figure}

Similar to \citet{hk17} we assume that water levels are in the interval spanned by half of the minimum and double of the maximum recorded water level, i.e. they are between 17.5 cm and 1650 cm, so the Box-Cox transforms of these values serve as lower and upper bounds for the truncated normal distribution used both in BMA and EMOS modelling.

The generation of the hydrological ensemble forecast described in Section \ref{sec:sec3} induces a natural grouping of the ensemble members. One contains just the forecast based on the ECMWF HRES, the other 51 member group corresponds to the ECMWF EPS, whereas forecasts based on COSMO LEPS and NCEP GEFS ensemble weather forecasts form two other groups of sizes 16 and 11, respectively. Hence, Box-Cox transformed water level forecasts are calibrated using the truncated normal BMA model for exchangeable ensemble members specified by \eqref{eq:genBMAex} and \eqref{eq:tnormPDF} and truncated normal EMOS given by \eqref{eq:emos} and \eqref{eq:emosEx} with \ $K=4$ \ and \ $M_1=1, \ M_2=51, \ M_3=16, \ M_4=11$. \ This means that for BMA modelling 12, whereas for finding the EMOS predictive distribution 7 free parameters have to be estimated. To ensure a reasonably stable parameter estimation we use a rolling training period of length 100 days and consider one day ahead calibration for all lead times. Thus, BMA and EMOS models are verified on the period 10 April 2008 -- 31 December 2015 (2822 calendar days).

While BMA and EMOS models are fit to Box-Cox transformed values, to ensure comparability we provide verification scores for the original forecasts and observations. This means that for quantile based scores (MAE, coverage, average width) before evaluating the score the inverse Box-Cox transformation is applied to the appropriate quantiles of the predictive distribution, whereas CRPS is calculated with the help of \eqref{eq:bcCRPS}.

\begin{figure}[!t]
    \epsfig{file=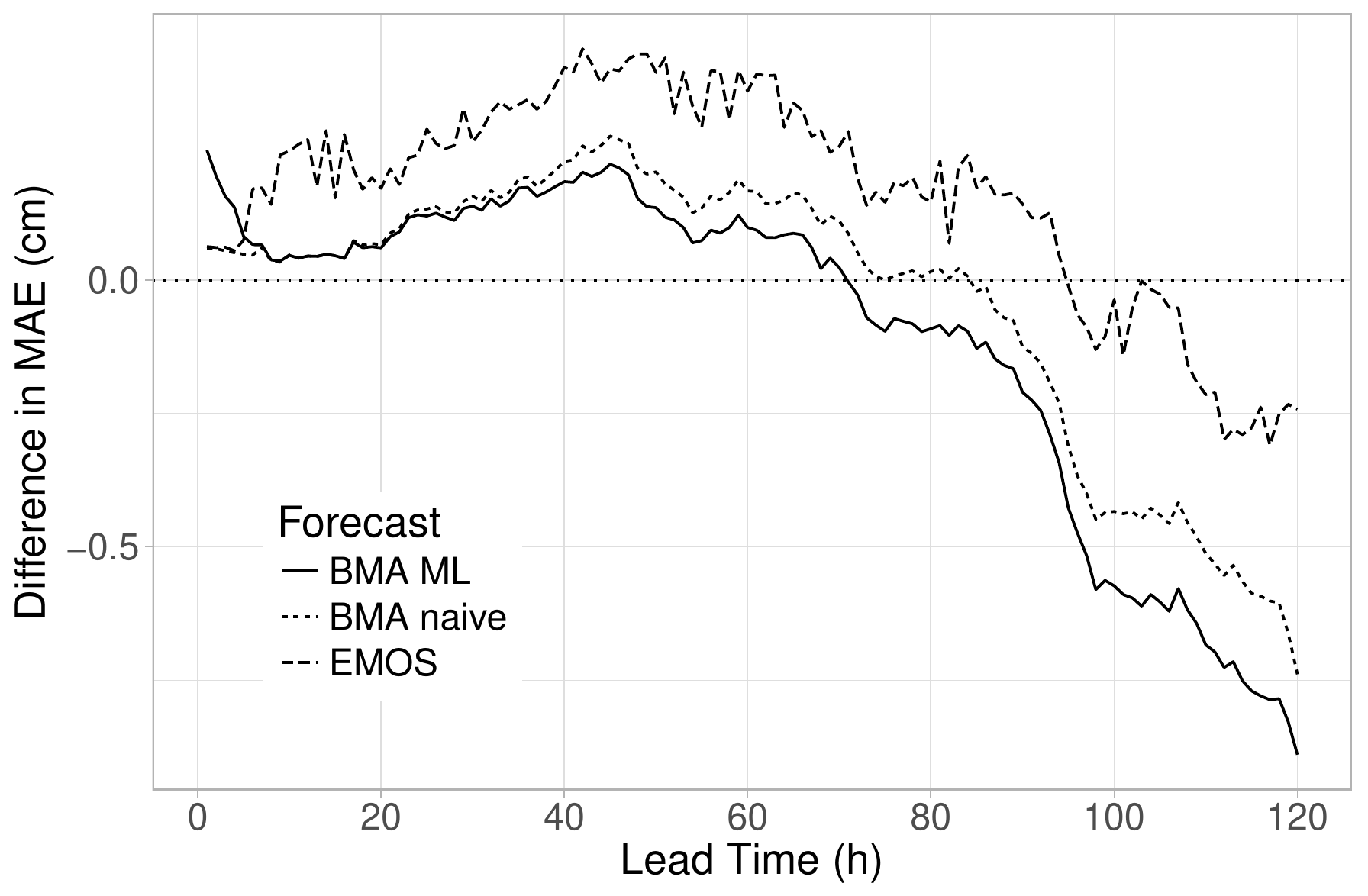, width=.49\textwidth} \ \
    \epsfig{file=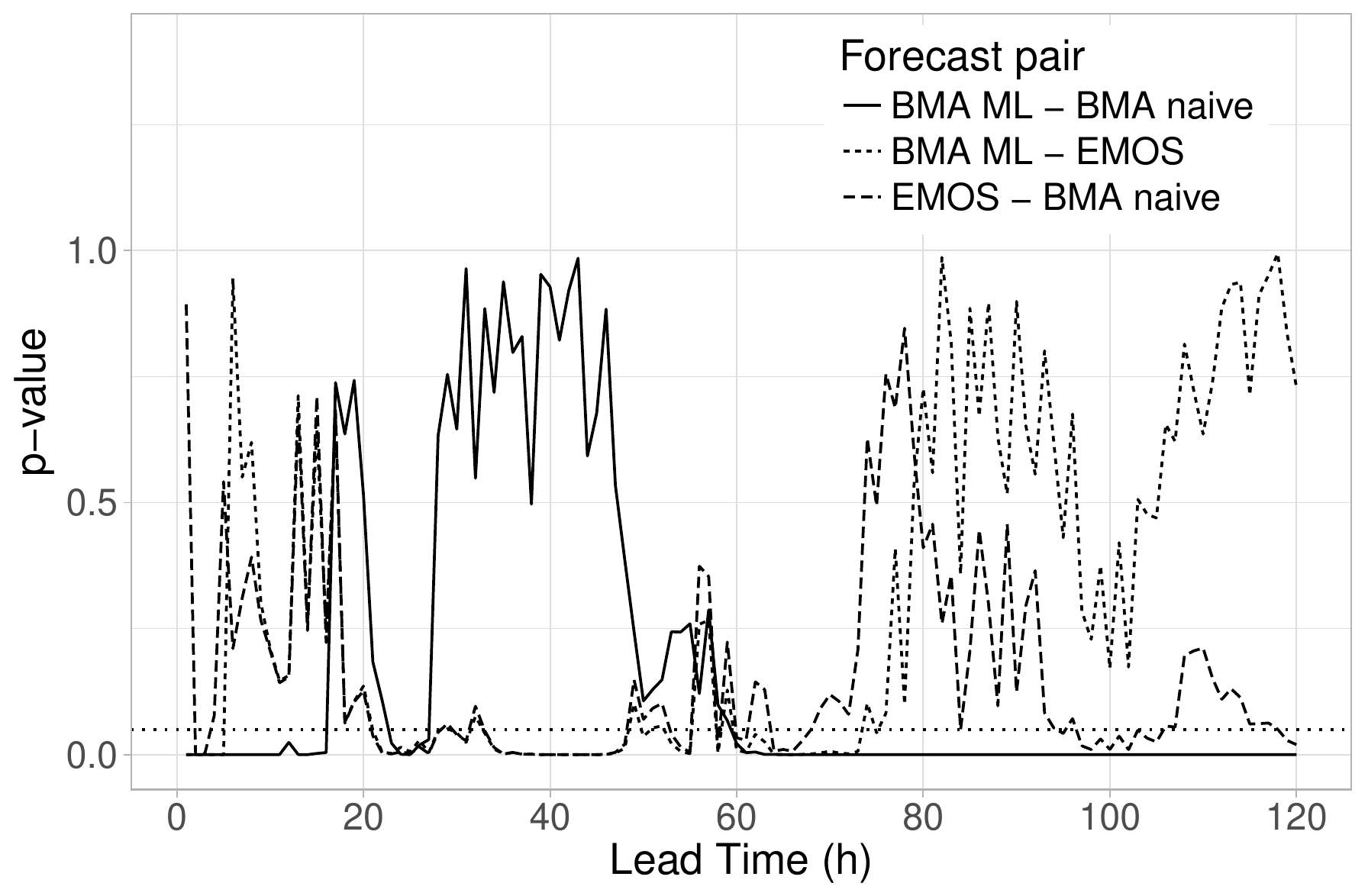, width=.49\textwidth}
    \centerline{ \hbox to 8.8 truecm{\scriptsize (a) \hfill (b)}}
\caption{Difference in MAE values from the raw ensemble (a) and $p$-values of DM tests for equality of MAE of the various post-processing approaches (b). Horizontal dotted lines indicate the reference raw ensemble (a) and a $5\,\%$ level of significance (b).}
\label{fig:mae}
\end{figure}

In Figure \ref{fig:crps}a the mean CRPS values of the different post-processing approaches and the raw ensemble are plotted as functions of the lead time. Note that compared to the raw ensemble all calibration approaches reduce the mean CRPS and the gap increases together with the lead time. The differences between the forecast skills are more pronounced in Figure \ref{fig:crps}b showing the CRPSS values with respect to the raw ensemble forecast. Note that all three presented methods have their maximal skill score at hour 9, for shorter lead times the increase is very fast and naive BMA shows the best predictive performance, whereas for longer lead times the pure ML BMA starts dominating. Obviously, larger lead times also result in larger forecast uncertainty which should be taken into account when one compares predictive performance. According to the results of DM tests for equal predictive performance naive BMA significantly outperforms the raw ensemble for all lead times and the same holds for the pure ML BMA except hour 1. In general, in terms of the mean CRPS the two BMA approaches differ significantly mainly for very short and long lead times, as can be observed on the graph of $p$-values displayed in Figure \ref{fig:crps}c. EMOS also significantly outperforms the raw ensemble for all lead times, and except for the first couple of hours underperforms the BMA approaches, as depicted in Figure \ref{fig:crps}d.

\begin{figure}[!t]
    \epsfig{file=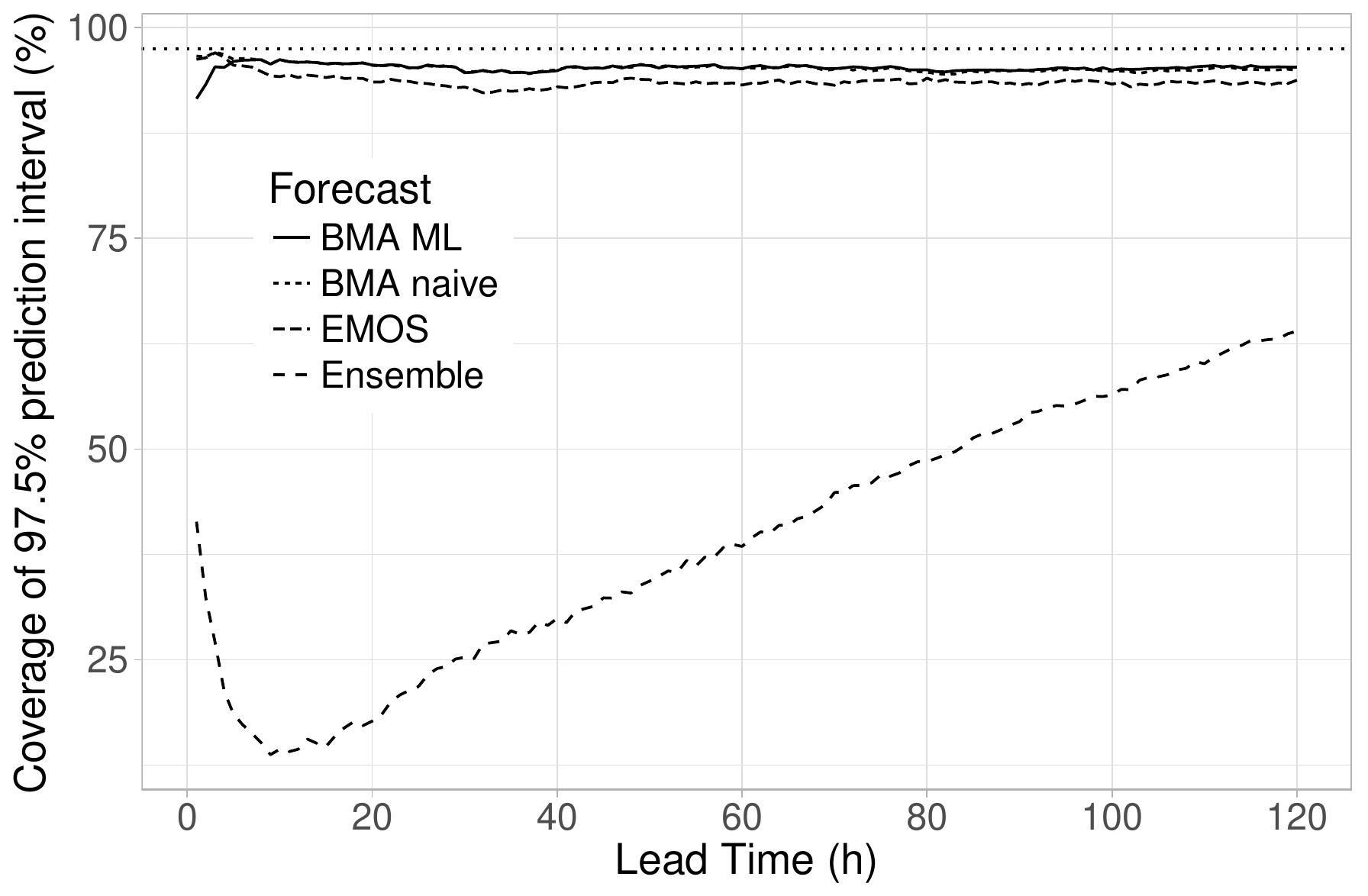, width=.49\textwidth} \ \
    \epsfig{file=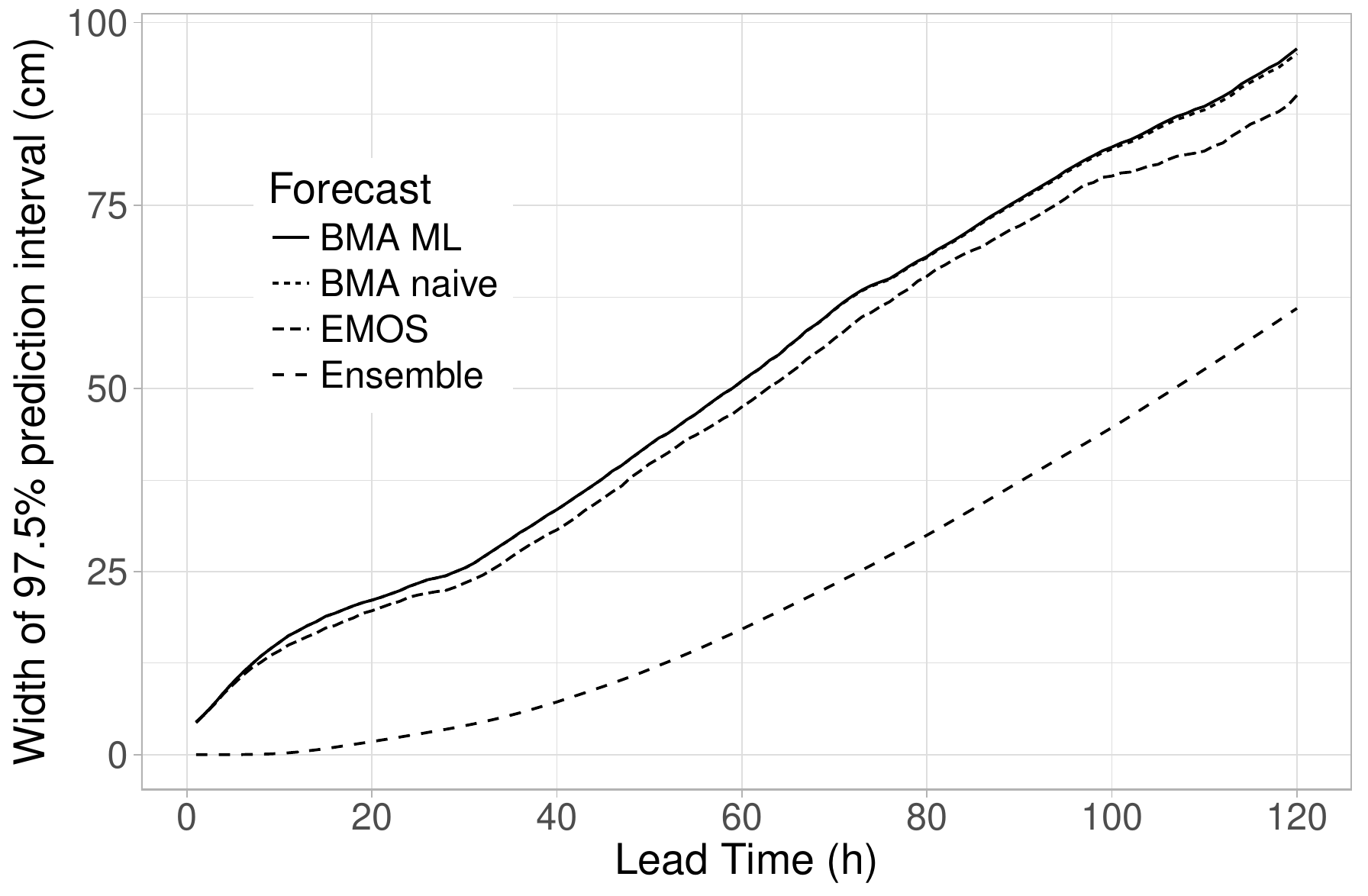, width=.49\textwidth}
    \centerline{ \hbox to 8.8 truecm{\scriptsize (a) \hfill (b)}}
\caption{Coverage (a) and average width (b) of nominal \ $97.5\,\%$ \ central prediction intervals.}
\label{fig:covaw}
\end{figure}

There is much less variety in the performance of BMA and EMOS calibrated medians in terms of the MAE. According to Figure \ref{fig:mae}a showing the difference in MAE with respect to the raw ensemble the pure ML BMA has the best forecast skill, however, even this approach underperforms the raw ensemble till hour 70. Note that DM tests for equality of MAE values indicate that all differences plotted in Figure \ref{fig:mae}a are significant, which will definitely not be the case if we compare the performance of the three post-processing methods, see the $p$-values of  Figure \ref{fig:mae}b.

\begin{figure}[!t]
  \begin{center}
  \epsfig{file=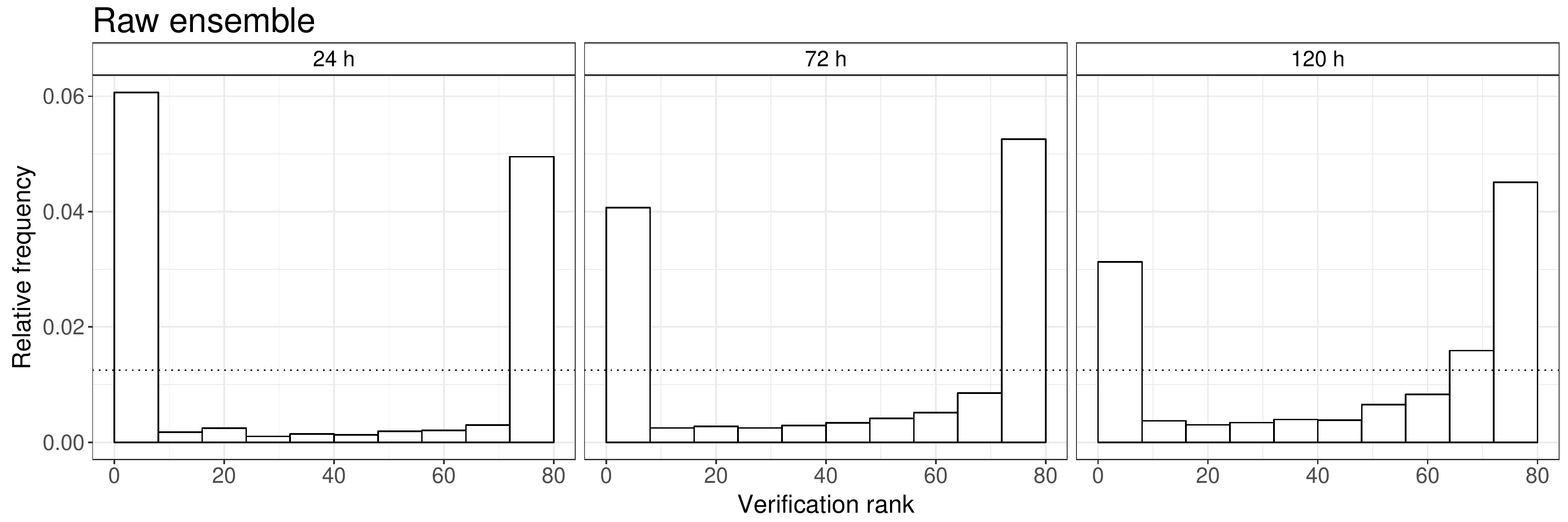, width=.91\textwidth}
  
  \epsfig{file=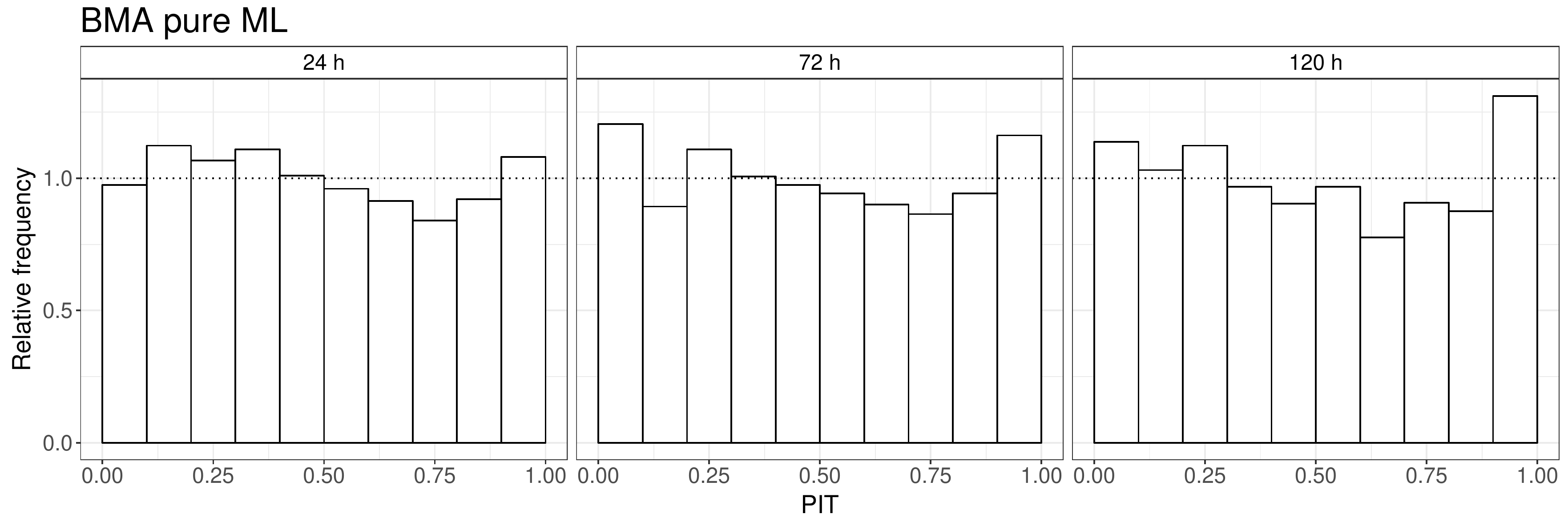, width=.91\textwidth}
  
  \epsfig{file=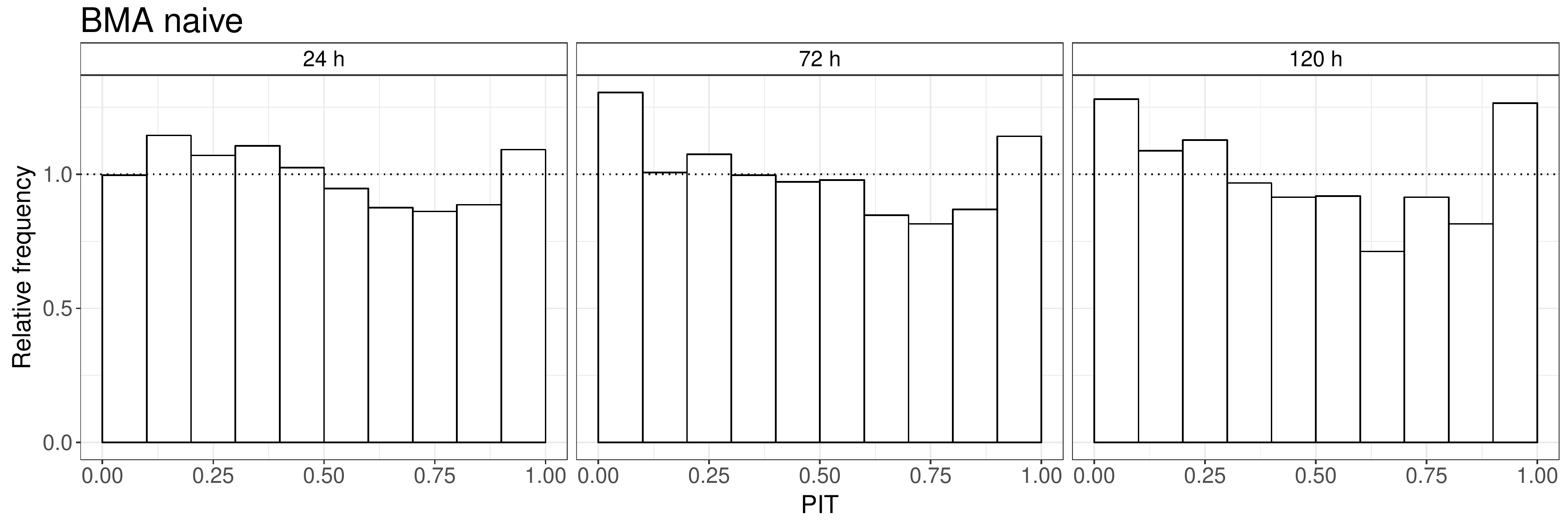, width=.91\textwidth}
  
  \epsfig{file=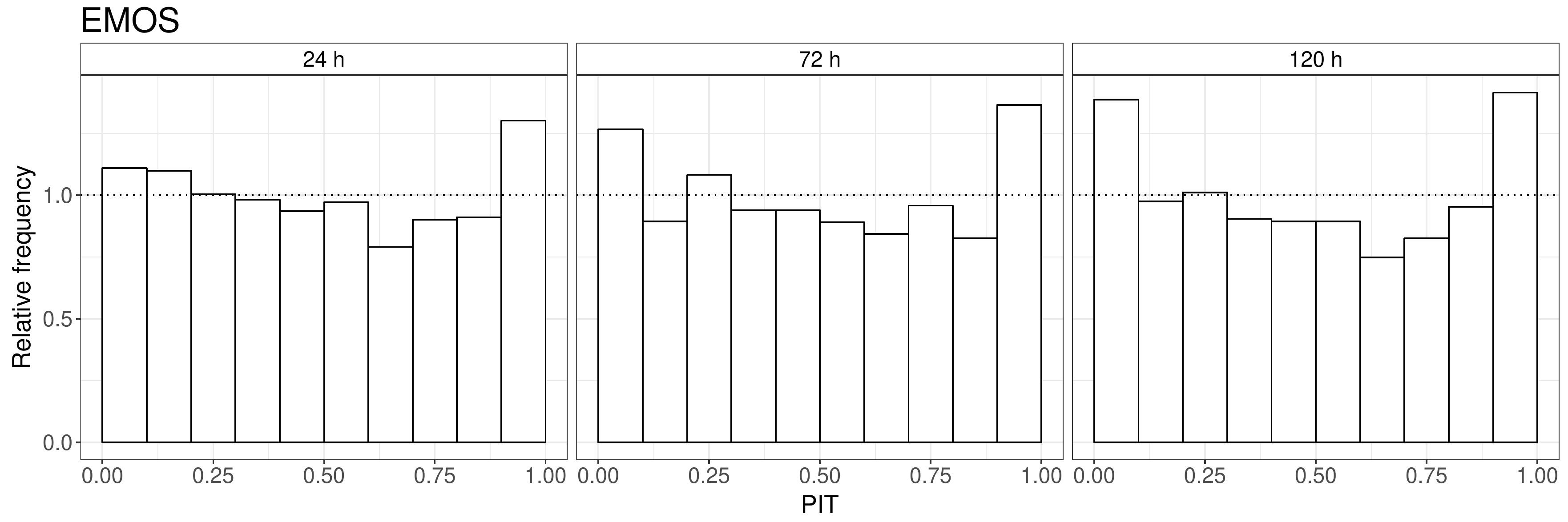, width=.91\textwidth}
\end{center}

\vskip -.3 cm
\caption{Verification rank histogram of the raw ensemble and PIT histograms of the BMA and EMOS post-processed forecasts for lead times 24, 72 and 120 hours.}
\label{fig:pit}
\end{figure}

\begin{figure}[t]
  \begin{center}
    \epsfig{file=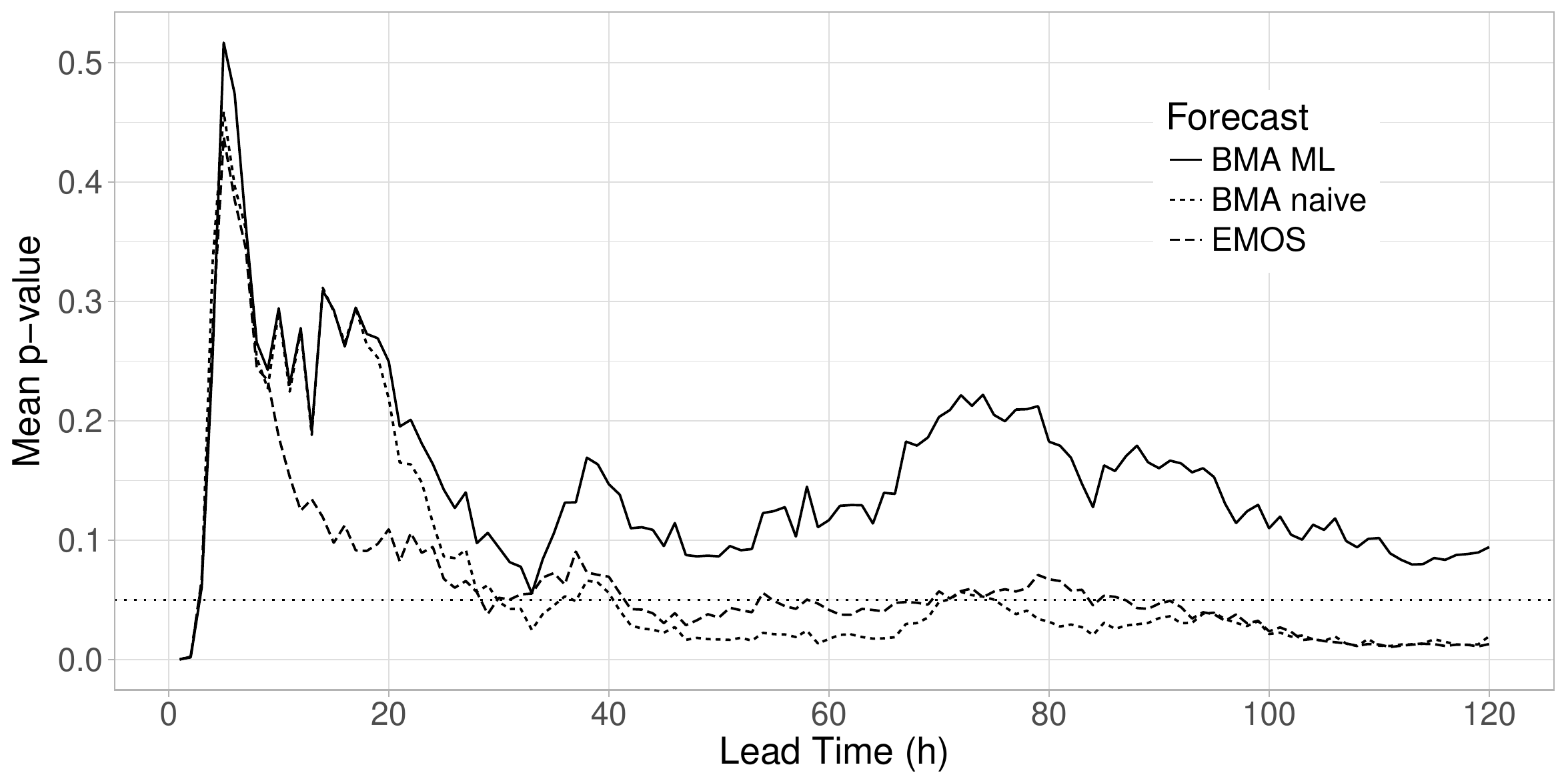, width=.7\textwidth}
    \end{center}
\caption{$p$-values of Kolmogorov-Smirnov tests for uniformity of PIT values. Average of 1000 random samples of sizes 1000 each.}
\label{fig:kstest}
\end{figure}

The positive effect of post-processing on calibration can be clearly observed on Figure \ref{fig:covaw}a showing the coverages of nominal \ $97.5\,\%$ \ central prediction intervals as functions of the lead time. All post-processing approaches for all lead times result in almost perfect coverage, whereas the coverage of the raw ensemble is much lover and strongly depends on the lead time. The coverage values of the two BMA approaches are almost identical and after hour 4 they are closer to the nominal value than those of the EMOS.
Finally, as depicted in  Figure \ref{fig:covaw}b, the raw ensemble produces the sharpest forecasts for all lead times, however, at the cost of being uncalibrated. This is fully in line with the  verification rank histograms of the raw ensemble and PIT histograms of post-processed forecasts for lead times 24, 72 and 120 hours plotted in Figure \ref{fig:pit}. All verification rank histograms are strongly U-shaped (and the same holds for other lead times, not reported), indicating that the raw ensemble is strongly underdispersive and requires post-processing. BMA and EMOS approaches significantly improve the statistical calibration of the forecast and result in more uniform PIT histograms, although for hour 120 naive BMA and EMOS still show a slight underdispersion. The Kolmogorov-Smirnov test accepts at a $5\,\%$ level the uniformity of the PIT values of pure ML BMA, naive BMA and EMOS for only 9 (5, 6, 7, 14, 17, 72, 75, 77, 79 h), 6 (4, 5, 6, 7, 14, 17 h) and 4 (5, 6, 7, 9 h) different lead times, respectively, however, this might be a consequence of the large sample size resulting in numerical problems \citep[see e.g.][]{bl15}. Thus, in order to get a better quantification of the differences in calibration, the mean $p$-values of 1000 random samples of PITs of sizes 1000 each, are calculated and plotted in Figure \ref{fig:kstest}. The corresponding points at lead times 24, 72 and 120 hours nicely reflect the shapes of the PIT histograms of Figure \ref{fig:pit} and clearly show the advantage of pure ML BMA post-processing.

\section{Conclusions}
\label{sec:sec5}
We introduce a new BMA model for calibrating Box-Cox transformed hydrological ensemble forecasts for water level, providing a predictive distribution which is a weighted mixture of doubly truncated normal distributions. The model with three different parameter estimation approaches is tested on the 79 member ensemble forecast of BfG for water level at gauge Kaub of river Rhine for 120 different lead times. Using the CRPS of probabilistic and MAE of median forecasts and coverage and average width of nominal central prediction intervals as verification measures, the forecast skill of the BMA model is compared to that of the state of the art EMOS model of \citet{hk17} and the raw ensemble. \\

Based on the results of the presented case study one can conclude that compared with the raw ensemble, post-processing always improves the calibration of probabilistic and accuracy of point forecasts. Further, BMA model utilizing pure ML for parameter estimation has the best predictive performance and, except very short lead times, the BMA approach significantly outperforms the EMOS calibration. A direct comparison of the CRPSS values obtained in this case study with those shown in Figure 4a of \citet{hk17} reveals that -- at least in the case of EMOS -- seasonal and analog based training periods considerably outperform the rolling window training periods used in this study. Hence, though out-of-scope of this study, a thorough comparison of the gains  in forecast skill and the (computational) costs of using a more complex post-processing method, e.g. BMA instead of EMOS with the gains and costs of using a more complex selection of the training periods, e.g. seasonal and analog based training periods for forecasts of water level, would be beneficial. 

Further, following the ideas of \citet{hlk15} one can combine the BMA calibrated forecasts corresponding to different lead times into temporally coherent multivariate predictions with the help of state of the art techniques such as e.g. the ensemble copula coupling \citep{stg13} or the Gaussian copula approach \citep{pg12}, however, these studies are also beyond the scope of the present paper.

\bigskip
\noindent
{\bf Acknowledgments.} \ The authors are grateful to the German Federal Office of Hydrology (BfG), and in particular Bastian Klein, for providing the data. The authors also thank Tilmann Gneiting for valuable comments and suggestions. Essential part of this work was made during the visit
of S\'andor Baran at the Heidelberg Institute of Theoretical Studies in the framework of the visiting scientist program.
S\'andor Baran was also supported by the J\'anos Bolyai Research Scholarship of the Hungarian Academy of Sciences and the National Research, Development and Innovation Office under Grant No. NN125679. He is grateful to Sebastian Lerch for his useful suggestions, remarks and helps with the EMOS code.
S\'andor Baran and Stephan Hemri also acknowledge the support of the
Deutsche Forschungsgemeinschaft (DFG) Grant No. MO 3394/1-1 
``Statistical postprocessing of ensemble forecasts for various weather quantities''.

\end{document}